\documentclass[aps,pre,letterpaper,twocolumn,superscriptaddress,floatfix]{revtex4-2}

\usepackage{graphicx,psfrag,amsmath,amssymb,amsfonts,bbm,latexsym}
\usepackage{color,xcolor,bm,braket,times,physics}
\usepackage{mathptmx,etoolbox,threeparttable}
\usepackage[dvipsnames]{xcolor}

\usepackage[utf8]{inputenc} \usepackage[T1]{fontenc}
\usepackage[english]{babel} \usepackage{slashed}

\begin{document}
	
\title{Maximization of nuclear fusion reactivity via optimal energy
  distributions}

\author{Roberto Onofrio} \thanks{Contact author: onofrior@gmail.com}
\affiliation{\mbox{Department of Physics and Astronomy, Dartmouth
    College, 6127 Wilder Laboratory, Hanover, New Hampshire 03755, USA}}
		
\author{Kamel Ourabah} \thanks{Contact author: kam.ourabah@gmail.com}
\affiliation{\mbox{Faculty of Physics, University of Bab-Ezzouar,
    USTHB, Boite Postale 32, El Alia, 16111, Algiers, Algeria}}

\author{Carlo Presilla} \thanks{Contact author: carlo.presilla@uniroma1.it}
\affiliation{\mbox{Dipartimento di Matematica, Sapienza Universit\`a
    di Roma, Piazzale Aldo Moro 2, 00185 Rome, Italy }}

\begin{abstract}
  We investigate the optimization of nuclear fusion reactivity
  relative to the choice of energy probability distribution.  A single
  Maxwell-Boltzmann (MB) distribution is found to be optimal in the
  region where reactivity is a concave function of the average
  available kinetic energy, specifically near the reactivity
  peak. Conversely, in regions where the reactivity curve is convex,
  optimization is achieved through a bimodal MB distribution. We then
  discuss our results in the context of three classes of generalized
  energy distributions (superstatistics), which are modeled as a
  continuum of MB distributions.  As expected, in all cases examined,
  the reactivity gains provided by superstatistics are bounded by the
  proposed optimal distribution.  Finally, we apply these findings to
  deuterium-deuterium and deuterium-tritium fusion processes using
  empirical parametrizations of their respective cross sections.  For
  both reactions, employing optimal bimodal MB distributions at low
  average energies -- of particular interest for magnetically confined
  plasmas -- yields a reactivity enhancement of approximately a factor
  of 2.
\end{abstract}

\maketitle
\date{\today}
\maketitle

\section{Introduction}
The production of non intermittent energy with low levels of air and
water pollution is an important goal of science and technology in the
first half of the 21st century. In this context, the most promising
approach seems to be nuclear fusion, with the potential to
significantly improve desirable features with respect to nuclear
fission, the latter being a source of energy production since the
1950s.  In addition, nuclear fusion studies play a crucial role in
understanding complex nonlinear systems in general, and plays a
crucial role in a variety of processes in astrophysics, most notably
stellar evolution and primordial nucleosynthesis. Efforts are ongoing
to reach break even in the most promising case of deuterium-tritium
(DT) fusion \cite{ITER1,ITER2}, in spite of the complications related
to the need to produce and maintain enough levels of radioactive
tritium. Next to this, deuterium-deuterium (DD) fusion processes have
the advantage, with respect to DT, of widespread availability in
water, a greater share of the energy self-igniting the reaction, and
less damage induced by the emitted neutrons due to their lower energy,
which also implies lower levels of environmental radioactivity. The
biggest bottleneck for DD fusion is the fusion cross section which
peaks at a value 2 orders of magnitude lower than the corresponding
peak for DT fusion, with its peak also occurring at much higher
temperature.

While current efforts focused on DT fusion using magnetic confinement
aim at increasing plasma volume or its density, it may also be worth
discussing a possible path in which new physical effects are
identified to catalyze DD fusion, hopefully recovering the 2 orders of
magnitude gap in its cross section by boosting fusion
reactivity~\cite{Onofrio}. With this goal in mind, some assumptions
usually used in the evaluation of the dynamics of fusion processes
have been scrutinized. For instance, the dependence of tunneling
processes on the use of plane waves and Gaussian wave packets for the
colliding ions in a thermonuclear reaction was analyzed
\cite{OnofrioPresilla}. There are no ideal plane waves in nature, and
more specifically this approximation is rather questionable in the
case of confined gases because of both the size of the confinement of
each individual ion and the presence of ion-ion and ion-electron
scattering limiting the mean free path.  In order to discriminate
discrepancies due to the use of the Wentzel-Kramers-Brillouin
approximation and its variants, the discussion has been framed within
an exactly solvable potential~\cite{GWS}. Sensible differences in the
tunneling coefficient and reactivities have been observed depending on
the width of the Gaussian wave packet, as well as the temperature and
the effective mass of the scatterers. This result is also relevant for
the parametrization of fusion reaction via the Gamow factor typically
used in astrophysics \cite{Hwang}.

Furthermore, the assumption of an ideal thermal equilibrium related to
Maxwell-Boltzmann (MB) distributions has been repeatedly challenged
\cite{Harvey,Nath,Majumdar,Kolmes,Li,Onofrio,Xie1,Xie2,Kong,Squarer,Coppi,Ye,Nicolas}. Indeed,
maximizing fusion reactivity also implies the discussion of potential
advantages in using energy distributions in which the high energy tail
does not fall quickly at high energy like in the case of MB
distributions due to their asymptotic exponential dependence on
energy, in spite of the possible limitations due to the replenishing
of the high energy tail if it is operated in a continuous
mode~\cite{Rider}. In particular, attention has been focused on energy
probability distributions representing generalizations of Lorentzian
distributions, also named kappa (or $\kappa$) distributions
\cite{Livadiotis1}.  These energy distributions were empirically
discovered by fitting the data of electrons in space plasmas and then
gained a more foundational ground in the framework of non extensive
statistical mechanics~\cite{Tsallis,Livadiotis2}. Some examples of
reactivity gain using $\kappa$-distributions with respect to their MB
counterparts were discussed in Ref.~\cite{Squarer}, including the case
of Gaussian states. In the same contribution, the robustness of the
results with respect to the choice of the potential barrier was also
scrutinized.

Here we extend the considerations reported in Ref.~\cite{Squarer} to
identify general conditions for the energy distribution required to
optimize reactivity. Rather than assuming a specific non-Maxwellian
distribution, we consider a general picture in which the system is
described as a continuous mixture of MB distributions at different
temperatures. This approach is well suited to systems that are locally
in equilibrium but globally out of equilibrium. In such conditions,
the system can be partitioned into small cells, each relaxing rapidly
to a local MB distribution, while at larger scales the inverse
temperature $\beta$ fluctuates from cell to cell according to the
distribution $f(\beta)$. This description has proven highly successful
in capturing observed distributions across a wide range of
physical~\cite{Sattin,stur2,stur3,Wilk,Ourabah1,OurabahPRE} and non
physical systems~\cite{Market,air,nature}, which typically decay more
slowly than the MB distribution, thereby opening the possibility of
identifying temperature fluctuations that optimally enhance
reactivities.

This paper is organized as follows. In Sec. II, we recall the
formalism introduced in previous work \cite{OnofrioPresilla,Squarer}
to incorporate generalized kinematic conditions for both the energy
distribution of the particle ensemble and the momentum distribution of
a single particle with a given energy. We also present the expression
for the most general ensemble distribution as a continuous mixture of
MB distributions with a normalized weight function $f(\beta)$. In
Sec. II A, we establish the conditions under which $f(\beta)$
optimizes reactivity. This optimization is straightforward when the
target reactivity is globally concave; applying Jensen's
inequality~\cite{Jensen} demonstrates that the standard MB energy
distribution is the optimal solution. However, physical constraints
dictate that the reactivity cannot be concave everywhere, as it
rapidly drops to zero at both low and high energies. Consequently, a
more refined analysis is required, which is detailed in Section II B.
Section III discusses the relationship between our optimization
results and three classes of generalized energy distributions known as
superstatistics. In Sec. IV, the results of Secs. II and III are
applied to DT and DD fusion processes using the parametrization of the
cross section data described in Ref.~\cite{Bosch}, highlighting the
advantages of employing the optimized distribution. Finally,
concluding remarks on the applicability of our findings to optimizing
fusion processes in magnetically confined plasmas are presented in
Sec. V.

\section{Reactivity for generalized kinematic conditions}

We start by summarizing results already discussed in
Ref.~\cite{OnofrioPresilla,Squarer} for reactivities evaluated using
scattering of ions and specifying the notation. The discussion is
focused on, but not limited to, the simplest one-dimensional case for
which we used analytical expressions for tunneling coefficients. The
generalization to the three-dimensional case is, formally at least,
straightforward.

The mutual scattering of ions at a characteristic momentum $\hbar K$
is well described by a stationary wave function $\psi_K(x)$, which is
shaped by the mutual nuclear tunneling potential. While the standard
approach involves plane waves, more general situations can arise and
often offer clearer physical insight. For instance, $\psi_K(x)$ can be
represented by a wave packet, such as the Gaussian wave packet
considered in Ref.~\cite{OnofrioPresilla}.  It is convenient to work
in the reciprocal space $k$ by writing the ionic wave function in
terms of its Fourier transform $\varphi_K(k)$
\begin{align}
  \psi_K(x) =
  \frac{1}{\sqrt{2\pi}}
  \int_{-\infty}^{+\infty} \varphi_K(k) e^{i k x} dk, 
\end{align}
which shows that $\psi_K(x)$ is a superposition of plane waves with
wave vectors $k\in (-\infty,+\infty)$ distributed according to the
probability density $P_K(k) = \vert \varphi_K(k)\vert^2$.  With the
choice $\psi_K(x)=[2/(\pi\xi^2)]^{\frac14} e^{-(x-x0)^2/\xi^2+iKx}$,
we have
\begin{align}
  P_K(k)=\frac{\xi}{\sqrt{2\pi}} e^{-\xi^2(k-K)^2/2}.
  \label{PkK}
\end{align}
Note that these Gaussian wave packets have mean momentum
$\langle \hbar k \rangle = \hbar K$ and mean kinetic energy
$\langle \hbar^2k^2/(2m) \rangle = \hbar^2 (K^2+\xi^{-2})/(2m)$, the
case of a pure plane wave with wave vector $K$ being recovered for
$\xi \to \infty$ when $P_K(k)\to\delta(h-K)$.
 
As stated, the above description applies to the case of monoenergetic
ions with characteristic momentum $\hbar K$, which may occur in ion
beams.  In reality, especially for thermonuclear plasmas, the wave
vector $K$ should be considered a real parameter drawn from a
statistical distribution $w(K)$ corresponding to some classical ionic
setup.  Thus, regarding $P_K(k)w(K)$ as the joint probability
distribution of the wave vectors $k$ and $K$ allows us to consider
fusion processes taking place in a statistical ensemble of nuclei
represented by Gaussian wave packets with an arbitrary distribution of
their mean momenta.

In the context of thermonuclear fusion, ions are usually assumed to
follow a MB energy distribution. Accordingly, for $w(K)$ we take the
following MB probability density, normalized to unity for
$K\in(-\infty,+\infty)$:
\begin{align}
  w_\mathrm{MB}(K;\beta) =
  \left(\frac{\beta}{\pi} \frac{\hbar^2}{2m} \right)^{1/2}
  e^{-\beta \hbar^2 K^2/(2m)}.
  \label{wMB}
\end{align}
As usual, $\beta=(k_B T)^{-1}$ is the inverse temperature
characterizing the MB distribution and $m$ the reduced mass of the two
ions $a$ and $b$ involved in the fusion reaction, namely,
$m=m_a m_b/(m_a+m_b)$.

An intensive measure of the reaction rates is provided by the
reactivity $\langle \sigma(E) v(E) \rangle$, where
$\langle \cdot \rangle$ denotes the average over the statistical
distribution of the reactants, $\sigma(E)$ is the cross section of the
process, and $v(E)$ is the velocity of the reduced ion.  In one
dimension and for nuclei at energy $E=m v^2/2=\hbar^2 k^2/(2m)$, the
cross section is~\cite{Satchler} $\sigma(E) = T(E) \pi / k^2$ , and
\begin{align}
  \sigma(E) v(E) =
  \frac{\pi \hbar^2}{\sqrt{2 m^3}} \frac{T(E)}{\sqrt{E}}.
\end{align}
For a statistical ensemble of nuclei as described above in terms of
the wave vectors $k$ and $K$, the reactivity must be averaged over the
joint probability distribution $P_K(k)w(K)$:
\begin{align}
  \langle \sigma v \rangle
  =\frac{\pi \hbar^2}{\sqrt{2 m^3}} \int_{-\infty}^{+\infty} dk
  \int_{-\infty}^{+\infty} dK ~
  \frac{T(E(k))}{\sqrt{E(k)}} ~  P_K(k) w(K),
  \label{reactivity}       
\end{align}
where $E(k) = \hbar^2 k^2/(2m)$.

In the case of Gaussian wave packets with mean momenta MB distributed
at inverse temperature $\beta$, that is, for $P_K(k) w(K)$ given
by~\eqref{PkK} and~\eqref{wMB}, the integral over $K$
in~\eqref{reactivity} can be performed by reducing the reactivity to
the simpler expression
\begin{align}
  \langle \sigma v \rangle_\mathrm{MB}(\beta)
  = \sqrt{\frac{\pi}{2}}
  \frac{\hbar \xi_{\mathrm{eff}}}{m}
  \int_{-\infty}^{+\infty} dk ~ \frac{T(E(k))}{k} ~ 
  e^{-\xi_{\mathrm{eff}}^2k^2/2},
  \label{reactivity-MB}
\end{align}
where $\xi_{\mathrm{eff}}^{-2} = \xi^{-2}+m/(\beta \hbar^2)$.  In the
limit $\xi^2 \gg \beta \hbar^2/m$, i.e., for wave packets
approximating a plane wave, the effective positional spreading becomes
the De Broglie thermal wavelength
$\xi_{\mathrm{eff}} \simeq (\beta \hbar^2/m)^{1/2}$.  In the opposite
limit of spatially narrow wave packets, $\xi^2 \ll \beta \hbar^2/m$,
we have $\xi_{\mathrm{eff}} \simeq \xi$.  Note that, even if we
prepare the ions in states with positional variance of quantum nature,
at high enough temperatures the relevant length scale below which
quantum coherence is maintained no longer depends on the initial
preparation, a result confirming earlier
discussions~\cite{Alterman,Chenu}.

A natural question is how the averaged reactivity~\eqref{reactivity}
is affected by the distribution $w(K)$.  For a MB distribution at a
fixed inverse temperature $\beta_0$, specifically,
$w(K)=w_\mathrm{MB}(K;\beta_0)$, we already obtained the
result~\eqref{reactivity-MB} with
$\xi_{\mathrm{eff}}^{-2} = \xi^{-2}+m/(\beta_0 \hbar^2)$. Here, we
explore the most general case of a $K$-distribution derived as a
continuous mixture of MB distributions at any possible inverse
temperature $\beta$
\begin{align}
  w_f(K) = \int_0^\infty d\beta\, f(\beta;\beta_0) ~ w_\mathrm{MB}(K;\beta).
  \label{w-f}
\end{align}
This distribution is fully characterized by a normalized weight
function $f(\beta)$, with $\beta \geq 0$.  More precisely, we specify
that $f(\beta)=f(\beta;\beta_0)$ also includes a parametric dependence
on $\beta_0 \geq 0$ to fairly compare with the MB distribution at that
inverse temperature.  In fact, while for
$f(\beta;\beta_0)=\delta(\beta-\beta_0)$ we have
$w_f(K)=w_\mathrm{MB}(K;\beta_0)$, for $f(\beta;\beta_0)$ to be a
smooth function of $\beta$ we require that the mean energy of $w_f(K)$
is equal to the mean energy of $w_\mathrm{MB}(K;\beta_0)$.  Recalling
that the mean energy of a MB distribution at inverse temperature
$\beta$ amounts to $1/(2\beta)$, we conclude that $f(\beta;\beta_0)$
can be arbitrary as long as subject to the constraints
\begin{align}
  \left\{
  \begin{aligned}
    &f(\beta;\beta_0)  \geq 0 ,
    \\
    &\int_0^\infty d\beta ~ f(\beta;\beta_0) = 1 ,
    \\
    &\int_0^\infty d\beta ~ f(\beta;\beta_0) ~ \frac{1}{2\beta}
      = \frac{1}{2\beta_0} .
  \end{aligned}
  \right. 
  \label{constraints}
\end{align}

\subsection{Optimization for a concave reactivity}

According to Eq.~\eqref{reactivity} and using the result
\eqref{reactivity-MB}, the average reactivity associated with the
distribution~\eqref{w-f} reduces to the following functional of
$f(\beta;\beta_0)$
\begin{align}
  \langle \sigma v \rangle_f(\beta_0) 
  &=\frac{\pi \hbar^2}{\sqrt{2 m^3}} \int_{-\infty}^{+\infty} dk
    \int_{-\infty}^{+\infty} dK ~
    \frac{T(E(k))}{\sqrt{E(k)}} ~  P_K(k) w_f(K)
    \nonumber \\
  &= \int_0^\infty d\beta ~ f(\beta;\beta_0) ~
    \langle \sigma v \rangle_\mathrm{MB}(\beta) .
    \label{reactivity-f}       
\end{align}
We seek to maximize the value of $\langle \sigma v \rangle_f(\beta_0)$
with the constraints~\eqref{constraints} and compare the gain with
respect to the MB average reactivity
$\langle \sigma v \rangle_\mathrm{MB}(\beta_0)$ at the same average
energy $1/(2\beta_0)$.  In the following, we will use
$0 < \beta <\infty$ as the domain of $f(\beta;\beta_0)$ even if the
support for this function is physically restricted to a finite
interval $[\beta_\mathrm{min},\beta_\mathrm{max}]$.

Based on the above statements, we are dealing with a constrained
linear variational problem which is conveniently recast as a
constrained probabilistic functional problem, namely, finding the
probability density function (PDF) $f(\beta;\beta_0)$ such that
\begin{align}
  \left\{
  \begin{aligned}
    &\mathsf{E}_f [ \langle \sigma v \rangle_\mathrm{MB}(\beta) ]
      \mbox{ is maximum} ,
    \\
    &\mathsf{E}_f [ \beta^{-1} ] = \beta_0^{-1} ,
  \end{aligned}
  \right.
  \label{probabilistic1}
\end{align}
where
$\mathsf{E}_f(\cdot) = \int_0^\infty d\beta\, f(\beta;\beta_0)
(\cdot)$ is the expectation induced by the PDF $f(\beta;\beta_0)$.  It
is easier if we rewrite the second equation of~\eqref{probabilistic1}
as a condition involving a moment of first order. To this end, we
change the variable $\beta=u^{-1}$ and define
$p(u;u_0) = u^{-2} f(u^{-1};u_0^{-1})$, where $u_0^{-1}=\beta_0$.
Probability is conserved, $f(\beta;\beta_0) d\beta = -p(u;u_0) du$.
In terms of the new PDF $p(u;u_0)$ and the assigned function
$h(u)= \langle \sigma v \rangle_\mathrm{MB}(u^{-1})$, the problem
\eqref{probabilistic1} becomes, for a given value of $u_0$, the search
for the PDF $p(u;u_0)$ such that
\begin{align}
  \left\{
  \begin{aligned}
    &\mathsf{E}_{p} \left[ h(u) \right] \mbox{ is maximum} ,
    \\
    &\mathsf{E}_{p} \left[ u \right] = u_0 ,
  \end{aligned}
  \right.
  \label{probabilistic2}
\end{align}
where $\mathsf{E}_{p}(\cdot) = \int_0^\infty du\, p(u;u_0) (\cdot)$ is
the expectation induced by $p(u;u_0)$.  Note that $u$ ($u_0$) are
twice the mean energy of a MB at inverse temperature $\beta$
($\beta_0$) and $h(u)$ is nothing but the MB average reactivity as a
function of the thermal energy $u=k_B T=\beta^{-1}$.

Problem \eqref{probabilistic2} falls within the class of the so-called
generalized moment problems~\cite{GMP}, more exactly, the sharp bounds
on expectations~\cite{SBE1,SBE2} and its solution depends on the shape
of $h(u)$ and, of course, the value of the parameter $u_0$.

Let us assume that the domain of $h(u)$ is
$(u_\mathrm{min},u_\mathrm{max})$, possibly with $u_\mathrm{min}=0$
and $u_\mathrm{max}=\infty$.  If $h(u)$ is concave everywhere in its
domain, the following Jensen inequality then applies for any
$u \in (u_\mathrm{min},u_\mathrm{max})$:
\begin{align}
  \mathsf{E}_{p} \left[ h(u) \right]
  \leq h \left( \mathsf{E}_{p}[u] \right) = h(u_0).   
\end{align}
The max of $\mathsf{E}_{p} \left[ h(u) \right]$ is thus obtained by
concentrating the PDF at point $u_0$, namely,
\begin{align}
  p(u;u_0) = \delta(u - u_0).
\end{align}
Note that the constraint $\mathsf{E}_{p}[u]=u_0$ is satisfied and $p$
is correctly normalized.

\subsection{Reactivity with concave and convex regions}

In general, $h(u)$ is not concave everywhere and the solution of the
problem \eqref{probabilistic2} requires the introduction of its
concave envelope $\hat{h}(u)$, namely, the tightest concave function
lying above or on $h(u)$. It is easy to visualize $\hat{h}$ as a
rubber band stretched over $h$ (see Fig.~\ref{OptimalFig4} below).  In
fact, according to Carath\'eodory's theorem~\cite{Cara} applied to
moment space, any valid expectation vector
$\left( E[u],E[h(u)] \right)$ lies on the concave envelope of the
curve $\{ (u,h(u)):~u\in(u_\mathrm{min},u_\mathrm{max}) \}$ and can be
represented by a discrete probability measure with finite
support. Specifically, since the constraint is given here by a moment
of order 1 (the mean), the optimal PDF $p(u;u_0)$ is supported by at
most two points.

Let $\hat{h}(u)$ be the smallest continuous concave function such that
$\hat{h}(u)\geq h(u)$ for any $u \in (u_\mathrm{min},u_\mathrm{max})$.
This concave envelope is easily visualized in one dimension as the
rubber band connecting the points $(u_\mathrm{min},h(u_\mathrm{min}))$
and $(u_\mathrm{max},h(u_\mathrm{max}))$ stretched over the $h$
function's local maxima. Depending on the value of $u_0$ in
$(u_\mathrm{min},u_\mathrm{max})$, there are two different types of
solutions $p(u;u_0)$ of \eqref{probabilistic2}.

If $u_0$ is in a range such that $\hat{h}(u_0) = h(u_0)$, namely, in
the proximity of the maxima of $h(u)$ where this function is concave,
then we have
\begin{align}
  p(u;u_0) = \delta(u - u_0) ,
  \label{popt_concave}
\end{align}
the normalized distribution which, as we have already seen, maximizes
the expectation of $h(u)$ satisfying, at the same time, the constraint
on the average of $u$.

If $u_0$ is in a range, say $u_1 \leq u_0 \leq u_2$ with
$\hat{h}(u_1) = h(u_1)$, $\hat{h}(u_2) = h(u_2)$ and
$\hat{h}(u) > h(u)$ for $u_1 < u <u_2$, it is more convenient to
concentrate the probability at the endpoints $u_1$ and $u_2$.  To
explain this behavior, we start observing that
\begin{align}
  \mathsf{E}_{p} \left[ h(u) \right]
  \leq \mathsf{E}_{p}[\hat{h}(u)] .
\end{align}
Since $\hat{h}(u)$ is the function growing linearly from
$(u_1,h(u_1))$ to $(u_2,h(u_2))$, namely,
\begin{align}
  \hat{h}(u) = h(u_1) + [h(u_2)-h(u_1)] \frac{u-u_1}{u_2-u_1},
  \qquad u_1 \leq u \leq u_2 ,
\end{align}
for the linearity we have
$\mathsf{E}_p[\hat{h}(u)] = \hat{h}(\mathsf{E}[u]) = \hat{h}(u_0)$,
and we conclude that for any $u_0 \in [u_1,u_2]$
\begin{align}
  \mathsf{E}_{p} \left[ h(u) \right]
  \leq \hat{h}(u_0) =
  \frac{u_2-u_0}{u_2-u_1} h(u_1) + \frac{u_0-u_1}{u_2-u_1}  h(u_2) .
  \label{inequality_hath}
\end{align}
The max of $\mathsf{E}_{p} \left[ h(u) \right]$ is achieved when
expression \eqref{inequality_hath} holds with equality, i.e., when
$p(u;u_0)$ concentrates the mass at $u_1$ and $u_2$
\begin{align}
  p(u;u_0) = w_1 \delta(u-u_1) + w_2 \delta(u-u_2),
  \label{popt_convex}
\end{align}
with weights
\begin{align}
  \qquad w_1=\frac{u_2-u_0}{u_2-u_1},
  \qquad w_2=\frac{u_0-u_1}{u_2-u_1}.
  \label{w1}
\end{align}

The PDF~\eqref{popt_convex} is correctly normalized and satisfies the
constraint $\mathsf{E}_{p} [u] = u_0$.  It only remains to discuss the
nature of the endpoints $u_1$ and $u_2$.  Bearing in mind the rubber
band analogy, there are four possible cases.  We may have
$u_1=u_\textrm{min}$ and $u_2=u_\textrm{max}$ or the two cases in
which just one end point coincides with one end point of the domain
\begin{align}
  &u_1=u_\textrm{min}, \quad \mbox{ $u_2$ such that }
    h'(u_2) = \frac{h(u_2)-h(u_1)}{u_2-u_1} ,
    \label{points-left}
  \\
  &\mbox{$u_1$ such that }
    h'(u_1) = \frac{h(u_2)-h(u_1)}{u_2-u_1}, \quad u_2=u_\textrm{max},  
    \label{points-right}
\end{align}
or, last, $u_\textrm{min} <u_1 < u_2 < u_\textrm{max}$ determined by
the bitangent condition
\begin{align}
  h'(u_1) = \frac{h(u_2)-h(u_1)}{u_2-u_1} = h'(u_2).
\end{align}

Let us specialize the above general scheme to the case
$h(u)=\langle \sigma v \rangle_\mathrm{MB} (k_B T)$.  For physical
reasons, the MB reactivity is not concave
everywhere~\cite{OnofrioPresilla}, it must vanish faster than linearly
at small and large temperatures and reach a finite maximum value at an
intermediate temperature. It follows that $h(u)$ is a bounded
continuous function going from being convex to concave and then convex
again in the domain $(u_\mathrm{min},u_\mathrm{max})$.  We thus divide
this domain into three regions: the left region where
$\hat{h}(u) > h(u)$ with endpoints
\begin{align}
  \left\{
  \begin{aligned}
    & u_1^{(l)} = u_\mathrm{min} ,
    \\
    & u_2^{(l)} \mbox{ such that } h'(u_2^{(l)}) =
      \frac{ h(u_2^{(l)}) - h(u_1^{(l)}) }{ u_2^{(l)} - u_1^{(l)}} ,
  \end{aligned}
  \right. 
  \label{points_l}
\end{align}
the right region where $\hat{h}(u) > h(u)$ with endpoints
\begin{align}
  \left\{
  \begin{aligned}
    &u_1^{(r)} \mbox{ such that } h'(u_1^{(r)}) =
      \frac{ h(u_2^{(r)}) - h(u_1^{(r)}) }{ u_2^{(r)} - u_1^{(r)} } ,
    \\
    &u_2^{(r)} = u_\mathrm{max} ,
  \end{aligned}
  \right.
  \label{points_r}
\end{align}
and the remaining central region where $\hat{h}(u) = h(u)$.  The
concave envelope of $h(u)$ is the step-wise function
\begin{widetext}
  \begin{align}
    \hat{h}(u) =
    \left\{
    \begin{aligned}
      &\frac{ u_2^{(l)} - u }{ u_2^{(l)} - u_\mathrm{min} }
        h(u_\mathrm{min}) +
        \frac{ u - u_\mathrm{min} }{ u_2^{(l)} - u_\mathrm{min} }
        h(u_2^{(l)})
        \qquad &&u_\mathrm{min} < u < u_2^{(l)} ,
      \\
      &h(u) &&u_2^{(l)} \leq u \leq  u_1^{(r)} ,
      \\
      &\frac{ u_\mathrm{max} - u }{ u_\mathrm{max} - u_1^{(r)} }
        h(u_1^{(r)}) +
        \frac{ u -u_1^{(r)} }{ u_\mathrm{max} - u_1^{(r)} }
        h(u_\mathrm{max})
        \qquad &&u_1^{(r)} < u < u_\mathrm{max} .
    \end{aligned}
    \right.
    \label{concave_envelope}
  \end{align}
\end{widetext}
Note that in Eq.~\eqref{concave_envelope} we are left with the simple
task of determining the two points $u_2^{(l)}$ and $u_1^{(r)}$ from
the values of $h(u)$ and its first derivative $h'(u)$ which we can
consider known, at least numerically, at any point $u$.

In conclusion, according to \eqref{popt_concave} and
\eqref{popt_convex} the optimal PDF $p_\mathrm{opt}(u;u_0)$ solution
of problem \eqref{probabilistic2} is
\begin{widetext}
  \begin{align}
    p_\mathrm{opt}(u;u_0) =
    \left\{
    \begin{aligned}
      &\frac{ u_2^{(l)} - u_0 }{ u_2^{(l)} - u_\mathrm{min} }
        \delta(u-u_\mathrm{min}) +
        \frac{ u_0 - u_\mathrm{min} }{ u_2^{(l)} - u_\mathrm{min} }
        \delta(u-u_2^{(l)})
        \qquad &&u_\mathrm{min} < u_0 < u_2^{(l)} ,
      \\
      &\delta(u-u_0) &&u_2^{(l)} \leq u_0 \leq  u_1^{(r)} ,
      \\
      &\frac{ u_\mathrm{max} - u_0 }{ u_\mathrm{max} - u_1^{(r)} }
        \delta(u-u_1^{(r)}) +
        \frac{ u_0 -u_1^{(r)} }{ u_\mathrm{max} - u_1^{(r)} }
        \delta(u-u_\mathrm{max})
        \qquad &&u_1^{(r)} < u_0 < u_\mathrm{max} .
    \end{aligned}
    \right.
    \label{tildef_opt}
  \end{align}
\end{widetext}
Let $h_\mathrm{opt}(u_0)$ be the optimal average reactivity
corresponding to the above optimal PDF; we have
\begin{widetext}
  \begin{align}
    h_\mathrm{opt}(u_0) =
    \int_0^\infty du \ p_\mathrm{opt}(u;u_0) h(u)
    = \left\{
    \begin{aligned}
      &\frac{ u_2^{(l)} - u_0 }{ u_2^{(l)} - u_\mathrm{min} }
        h(u_\mathrm{min}) +
        \frac{ u_0 - u_\mathrm{min} }{ u_2^{(l)} - u_\mathrm{min} }
        h(u_2^{(l)})
        \qquad &&u_\mathrm{min} < u_0 < u_2^{(l)} ,
      \\
      &h(u_0) &&u_2^{(l)} \leq u_0 \leq  u_1^{(r)} ,
      \\
      &\frac{ u_\mathrm{max} - u_0 }{ u_\mathrm{max} - u_1^{(r)} }
        h(u_1^{(r)}) +
        \frac{ u_0 - u_1^{(r)} }{ u_\mathrm{max} - u_1^{(r)} }
        h(u_\mathrm{max}) 
        \qquad &&u_1^{(r)} < u_0 < u_\mathrm{max} .
    \end{aligned}         
    \right.
    \label{h_opt}
  \end{align}
\end{widetext}
By construction, the optimal average reactivity coincides with the
concave envelope function, $h_\mathrm{opt}(u_0) = \hat{h}(u_0)$.
Viewed as a function of $\beta_0$, the optimal average reactivity
$\langle\sigma v\rangle_\mathrm{opt}(\beta_0)$ is obtained from
$h_\mathrm{opt}(u_0)$ by applying the substitution
$u_0 \to \beta_0^{-1}$.

We close this section with an important observation. The optimization
procedure described above, despite being formulated for a
one-dimensional setup with a MB distribution in momentum space, yields
a completely general result. Indeed, the solution obtained for the
constrained probabilistic functional problem stated in
Eq.~\eqref{probabilistic1} remains valid for any arbitrary expression
of $\langle \sigma v \rangle_\mathrm{MB}(\beta)$, with the latter
playing the role of the assigned function for which we must find the
concave envelope.

\begin{figure*}[t]
  \includegraphics[width=0.33\textwidth, clip=true]{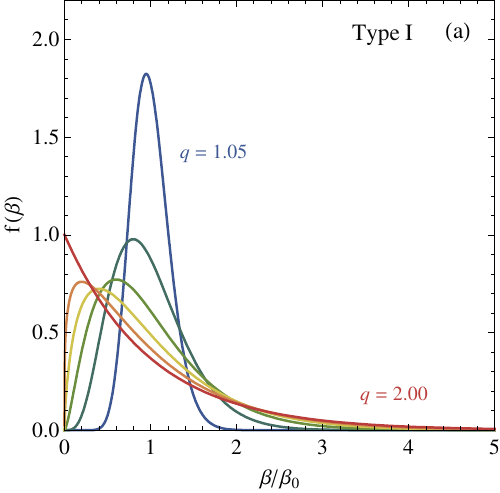}
  \includegraphics[width=0.33\textwidth, clip=true]{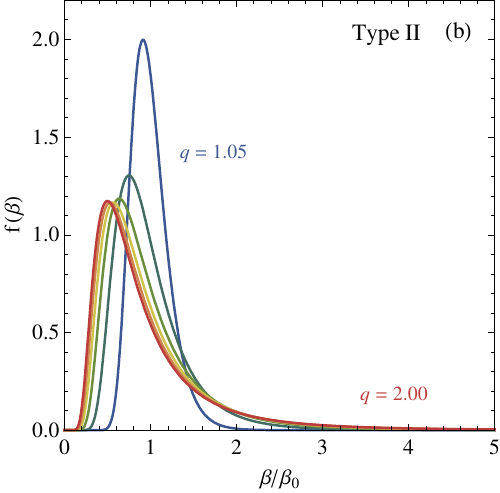}
  \includegraphics[width=0.33\textwidth, clip=true]{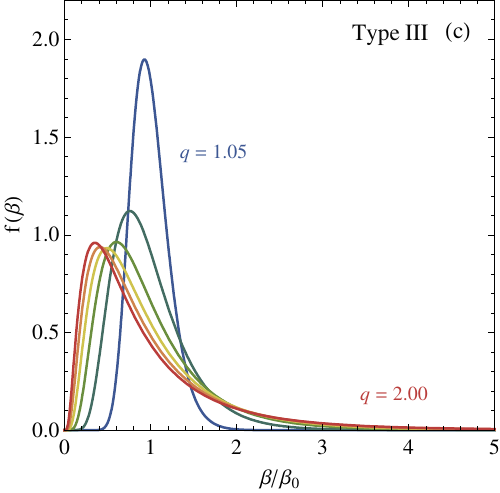}
  \caption{Probability distributions $f(\beta)$ for the three
    representative superstatistics, parametrized by $q$, shown for
    different values, starting from the leftmost peak at $q=2.0$ and
    going to the rightmost peak at $q=1.05$, with intermediate values
    of $q=1.20, 1.40, 1.60, 1.80$.  In the limit $q \to 1$, $f(\beta)$
    collapses to a Dirac $\delta$ distribution centered at $\beta_0$,
    and the corresponding energy distribution reduces to the MB form.}
  \label{OptimalFig1}
\end{figure*}

\section{Optimized energy distributions and superstatistics}

Superstatistics are a way to generalize MB distributions including the
case of effective temperatures spread out in a continuum
\cite{Beck1,Sattin,Beck2,Ourabah1,Ourabah2}. In this description, the
system is viewed as a superposition of local equilibrium states, each
characterized by a different value of $\beta$. The resulting energy
distribution is therefore obtained by averaging the MB distribution
over these inverse-temperature distribution $f(\beta)$—nonnegative and
normalized to unity once integrated in $\beta$—such that the energy
distribution of the $f$ superstatistics is written as
\begin{align}
  w_f(E)= \int_0^\infty d\beta\, f(\beta) w_\mathrm{MB}(E;\beta),
  \label{E-Superstatistics}
\end{align}
where
\begin{align}
  w_\mathrm{MB} (E;\beta) = \sqrt{\frac{\beta}{\pi E}} e^{-\beta E},
\end{align}
and
\begin{align}
  \quad\int_0^\infty dE\, w_\mathrm{MB} (E;\beta) = 1.
  \label{E-MB}
\end{align}
As usual, $\beta=1/(k_B T)$ indicates the inverse temperature, with
$k_B$ being the Boltzmann constant.

Sometimes it is preferable to work in terms of wave vectors $K$
instead of energies $E$. Assuming the dispersion relation of free
particles, $E=\hbar^2 K^2/(2m)$, the above relations for the MB
distribution become equivalent to Eq. (\ref{wMB}).  The $f$
superstatistics distribution expressed in terms of $K$ is then as
in~Eq.\eqref{w-f}
\begin{align}\label{wf}
  w_f(K)= \int_0^\infty d\beta\, f(\beta) w_\mathrm{MB}(K;\beta),
\end{align}
with $f(\beta) \geq 0$ and
\begin{align}
  \int_0^\infty d\beta\, f(\beta) = 1.
  \label{K-Superstatistics}
\end{align}
Note that, while the MB distribution is characterized by a given
inverse temperature $\beta$ connected to the average energy by
\begin{align}
  \int_0^\infty dE\, E w_\mathrm{MB} (E;\beta) =
  \int_{-\infty}^{+\infty} dK\, \frac{\hbar^2 K^2}{2m}
  w_\mathrm{MB} (K;\beta)= \frac{1}{2\beta},
  \label{average-energy-MB}
\end{align}
in the case of the $f$ superstatistics distributions we have only an
effective inverse temperature defined via the average energy
\begin{align}
  \int_0^\infty dE\, E w_f(E) =
  \int_{-\infty}^{+\infty} dK\, 
  \frac{\hbar^2 K^2}{2m} w_f(K)
  = \int_0^\infty d\beta\, f(\beta) \frac{1}{2\beta}.
  \label{average_energy-f}
\end{align}

We now focus on three representative examples of superstatistics: the
$\chi^2$ (type I), inverse $\chi^2$ (type II), and log-normal (type
III) classes.  These constitute the main universality classes
encountered across a wide range of physical systems and are supported
by substantial observational evidence. They can be traced back to
generic phenomenological mechanisms governing the stochastic
fluctuations of $\beta$~\cite{Beck2005} and may also be derived from a
maximum entropy framework under appropriate
constraints~\cite{Vander2008}.

In Fig.~\ref{OptimalFig1}, we present examples of the distributions
$f(\beta)$ for the three superstatistical classes, shown for different
values of
\begin{align}
  q:= \int_0^\infty d\beta\, f(\beta) \beta^2 \frac{1}{\beta_0^2}
  = \frac{\langle \beta^2 \rangle}{\beta_0^2} . 
  \label{q_common}
\end{align}
This parameter provides a common measure of inverse-temperature
fluctuations, since it is defined in the same way for all three
classes. Thus, comparing different distributions at the same value of
$q$ amounts to comparing them at the same degree of fluctuations. In
the limit $q\to1$, all three distributions collapse toward a Dirac
$\delta$ centered at $\beta_0$, recovering the MB limit.  As $q$
increases, the distributions broaden and their weight is redistributed
away from $\beta_0$ in a class-dependent way.  It is noticeable that
at least for the type-II and III superstatistics there is a leading
contribution at the inverse temperature corresponding to an MB
distribution, plus contributions with weight inversely proportional to
the distance from the peak. In the case of type-I superstatistics, the
peak of the distribution moves faster than the other two
superstatistics toward $\beta=0$, as shown more explicitly in Fig. 2
(a more detailed discussion is given below).  In general, many
statistical systems with relatively small number of particles, either
because they are of mesoscopic size or because they form small
subsystems with partial, only local thermal equilibrium, are better
described by a superstatistics taking into account temperature
fluctuations instead absent in a more macroscopic system manifesting
global equilibrium.

\begin{figure*}[t]
  \includegraphics[width=0.48\textwidth, clip=true]{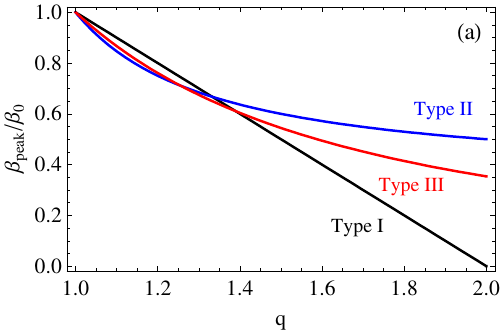}
  \hfill \includegraphics[width=0.48\textwidth,
  clip=true]{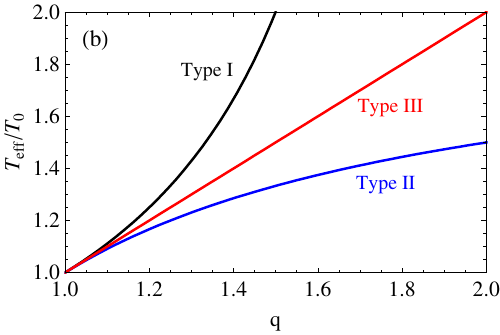}
  \caption{(a) Position of the maximum of $f(\beta)$,
    $\beta_{\mathrm{peak}}$, normalized to $\beta_0$, as a function of
    the universal parameter $q$ for the three superstatistics. (b)
    Effective temperature, normalized to the temperature $T_0$ of the
    MB energy distribution with the same average energy, versus $q$
    for the same three superstatistics.}
  \label{OptimalFig2}
\end{figure*}

A useful way to quantify how the weight of each distribution shifts
with increasing $q$ is to locate its maximum.  Denoting by
$\beta_{\mathrm{peak}}$ the value of $\beta$ at which $f(\beta)$ is
maximal, we obtain
\begin{align}
  \frac{\beta_{\mathrm{peak}}}{\beta_0}=2-q
\end{align}
for type I,
\begin{align}
  \frac{\beta_{\mathrm{peak}}}{\beta_0}=\frac{q}{3q-2}
\end{align}
for type II, and
\begin{align}
  \frac{\beta_{\mathrm{peak}}}{\beta_0}=q^{-3/2}
\end{align}
for type III. These three behaviors are shown in 
Fig.~\ref{OptimalFig2}(a). In all cases,
$\beta_{\mathrm{peak}}\to\beta_0$ as $q\to1$. As $q$ increases, the
peak moves toward smaller values of $\beta$, although with different
rates depending on the class. In particular, type I reaches
$\beta_{\mathrm{peak}}=0$ already at the limiting value $q=2$, while
type II tends to $\beta_0/3$ and type III tends to zero only
asymptotically as $q\to\infty$.

In what follows, for each of these classes, we evaluate the reactivity
under the condition of equal total energy and compare it with the
result obtained from the MB energy distribution with the same total
energy. We also restrict ourselves to the plane-wave case, for which
$P_K(k)=\delta(k-K)$ in Eq.~\ref{PkK}. Since the parameter
$q = \langle \beta^2 \rangle / \beta_0^2$ was already introduced
as a common measure of the strength of inverse-temperature
fluctuations, it will be used below to compare the three
superstatistical classes at the same degree of departure from the MB
limit.

For a more compact representation of the reactivity gains, we
introduce the dimensionless energy variable
$x := \beta_0 E(K)=\beta_0\hbar^2K^2/2m$ and the integral operator
\begin{equation}
  \mathcal{G}_{\beta_0}[F] :=
  \frac{\displaystyle
    \int_{-\infty}^{+\infty}dk\int_{-\infty}^{+\infty}dK\,
    \frac{T(E(k))}{\sqrt{E(k)}}~P_K(k)~F(x)}
  {\displaystyle
    \int_{-\infty}^{+\infty}dk\int_{-\infty}^{+\infty}dK\,
    \frac{T(E(k))}{\sqrt{E(k)}}~P_K(k)~e^{-x}
    \label{common_gain_operator}}.
\end{equation}
The dependence on the superstatistics class is then entirely contained
in the function $F(x)$ and its associated prefactor.

\subsection{Type I: $\bm{\chi^2}$ superstatistics}

A relevant example of superstatistics consists of a choice of a
two-parameter distribution $f=f_n(\beta; \beta_0)$ such that
\begin{align}
  f_n(\beta; \beta_0)= \frac{1}{\Gamma(n/2)}
  \left(\frac{n}{2\beta_0}\right)^{n/2} \beta^{n/2-1} e^{-n\beta/(2 \beta_0)},
  \label{Superstatistics1}
\end{align}
which represents a $\chi^2$ distribution of degree $n$ with generic
moment of order $k$
\begin{align}
  \langle \beta^k \rangle = \frac{\Gamma(n/2+k)}{\Gamma(n/2)}
  \left(\frac{2\beta_0}{n}\right)^k,
\end{align}
from which we can identify the average value of the inverse
temperature $\langle \beta \rangle=\beta_0$ and its variance
\begin{align}
  \Delta \beta^2=\langle \beta^2 \rangle - \langle \beta \rangle^2
  = \frac{2 \beta_0^2}{n}.
  \label{Variance}
\end{align}
It follows that for this superstatistics
\begin{align}
  q=\frac{\langle \beta^2\rangle}{\beta_0^2}=1+\frac{2}{n}. 
\end{align}
The resulting integral (\ref{wf}) is calculated analytically
\begin{widetext}
  \begin{align} \label{eq50} \int_0^{\infty} d \beta
    f_n^{\mathrm{\chi^2}} \left(\beta ; \beta_0\right) w_\mathrm{MB}(K,\beta)
    &= \frac{1} {\Gamma(n / 2)}\left( \frac{h^2}{2
      \pi m}\right)^{1 / 2}\left(\frac{n}{2 \beta_0}\right)^{n / 2}
      \int_0^{\infty} d \beta \beta^{\frac{n-1}{2}} \exp
      \left[-\left(\frac{\hbar^2 K^2}{2 m}+\frac{n}{2 \beta_0}\right)
      \beta\right]
      \nonumber \\
    &= \frac{\Gamma(\frac{n+1}{2})} {\Gamma(n/2)}
      \left( \frac{\beta_0 \hbar^2}{\pi m
      n}\right)^{1 / 2}\left(1+\frac{ \beta_0
      \hbar^2 K^2}{n m}\right)^{-(\frac{n+1}{2})}.
  \end{align}
\end{widetext}
Note that the $\chi^2$ distribution in Eq.~(\ref{Superstatistics1})
arises naturally when $\beta$ results from the superposition of many
independent random contributions, i.e., from an additive stochastic
process~\cite{Beck2005}. In this case, $\beta$ can be expressed as a
sum of squared Gaussian variables, leading, by virtue of the central
limit theorem, to a $\chi^2$ distribution, where $n$ represents the
number of degrees of freedom associated with the underlying
fluctuations. The corresponding energy distributions,
Eq.~(\ref{eq50}), exhibit power-law tails, making the $\chi^2$ class
particularly well suited to describe a wide range of nonequilibrium
systems characterized by enhanced high-energy tails, as commonly
observed in plasmas~\cite{azerty5,plasma1}, gravitational
systems~\cite{Astro1,galaxy,Our2020,Mahsa}, and various other physical
contexts~\cite{cold1,cold2,cold3,HE1,HE2}.

The reactivity gain obtained by comparing the $\chi^2$
superstatistical distribution with the MB distribution at the same
value of the inverse-temperature parameter $\beta_0$,
$G_n^{\chi^2}(\beta_0)=\langle\sigma v\rangle_f/\langle\sigma
v\rangle_\mathrm{MB}$, reads
\begin{equation}
  G_n^{\chi^2}(\beta_0)
  =\frac{\Gamma\left(\frac{n+1}{2}\right)}
  {\Gamma\left(\frac{n}{2}\right)}\sqrt{\frac{2}{n}}\,
  \mathcal{G}_{\beta_0}\!\left[
    \left(1+\frac{2x}{n}\right)^{-\frac{n+1}{2}}
  \right].
  \label{GainChi2Beta0}
\end{equation}

\begin{table*}
  \begin{threeparttable}
    \centering
    \begin{tabular}{|c|c|c|}
      \hline
      Statistics & $q$ parameter
      & $I_f(k,K)$ \mbox{defined in Eq.~(\ref{IntegralkK}}) \\
      \hline
      \hline
      $\chi^2$ & $1+2/n$
      & {\small{$\frac{\Gamma(\frac{q+1}{2(q-1)}}
        {\Gamma(\frac{1}{q-1})} \sqrt{\frac{q-1}{q-2}} 
        \left(\frac{\hbar^2 k^2}{2 m}\right)^{-1/2}
        T\left(\frac{\hbar^2 k^2}{2 m}\right)
        P_K(k)\left(1-\frac{(q-1) \hbar^2 K^2}
        {4(q-2) m \langle 
        E \rangle} \right)^{-\frac{q+1}{2(q-1)}}$}} \\
      \hline
      Inverse $\chi^2$ & $n/(n-2)$
      & {\small{$\frac{2^{-5/2 + \frac{1}{1-q}}
        \, \sqrt{2 + \frac{1}{q-1}}}
        {\Gamma\left(2 + \frac{1}{q-1}\right)}
        \left(\frac{\hbar^2 k^2}{2m}\right)^{-1/2} 
        T\left(\frac{\hbar^2 k^2}{2 m}\right)
        P_K(k) \left(\frac{(2q-1) \hbar^2 K^2}
        {(q-1)m \langle E \rangle}
        \right)^{\frac{3q-1}{2(q-1)}}
        \mathcal{K}_{\frac{3q-1}{2(q-1)}}
        \left( \sqrt{(2q-1)  \hbar^2 K^2/
        ((q-1) m \langle E \rangle)}\right)$}} \\
      \hline
      Log normal & $\exp(s^2)$
      & {\small{$\frac{1}{\sqrt{2 \pi \ln(q)}}  
        \left(\frac{\hbar^2 k^2}{2 m}\right)^{-1 / 2}
        T\left(\frac{\hbar^2 k^2}{2 m}\right) P_K(k)
        \int_0^{\infty} d \beta \sqrt{\frac{2
        \langle E \rangle}{\beta}}  
        \exp\left[-\frac{\beta \hbar^2 K^2}{2m}-
        \frac{(\ln(4 \beta \langle E \rangle /q))^2}
        {2 \ln(q)} \right]$}} \\ 
      \hline
    \end{tabular}
    \caption{Summary of the reactivities for the three classes of
      superstatistics. Shown are the relationships between the
      $q$ parameter and the index usually denoting each
      superstatistics, and the expressions for the integrand of the
      corresponding reactivities, expressed in terms of the average
      energy $\langle E \rangle$, as in Eq.~(\ref{IntegralkK}).}
    \label{Table}
  \end{threeparttable}
\end{table*}

\subsection{Type II: Inverse $\bm{\chi^2}$ superstatistics}

In the case of this second class of superstatistics, the temperature
$T= \beta^{-1}$ is $\chi^2$ distributed. That is,
\begin{align}
  f_n(T; T_0)= \frac{1}{\Gamma(n/2)} \left(\frac{n}{2T_0}\right)^{n/2}
  T^{n/2-1} e^{-n T/(2 T_0)},
\end{align}
where $T_0= \langle T \rangle$. Changing variables to $\beta$ (and
using $|dT/d \beta| = \beta^{-2}$), yields
\begin{align}
  {f_n}(\beta)=\frac{1}{\Gamma\left(\frac{n}{2}\right)}\left(\frac{n}{2
      T_0}\right)^{n / 2} \beta^{-n / 2-1} e^{-\frac{n }{2 \beta
      T_0}}.
  \label{f2T}
\end{align}
The mean of $\beta$ is
$\beta_0 \equiv \langle \beta \rangle= n/[(n-2) T_0]$ with
$n>2$. Using $\beta_0$ and redefining $n-2 \to n$, the inverse
$\chi^2$ distribution can be written as
\begin{align} {f_n}^{\mathrm{inv\chi^2}}(\beta)=\frac{\beta_0}
  {\Gamma\left(\frac{n}{2}\right)}
  \left(\frac{n\beta_0}{2}\right)^{n/2} \beta^{-n/2-2}
  e^{-\frac{n\beta_0}{2 \beta}},
  \label{superstatistics2}
\end{align}
with $n>0$. The variance in this case is
\begin{align}
  \Delta \beta^2
  =
  \langle \beta^2 \rangle-\langle \beta \rangle^2
  =
  \frac{2}{n-2}\beta_0^2
\end{align}
and
\begin{align}
  q=\frac{\langle \beta^2\rangle}{\beta_0^2}
  =
  \frac{n}{n-2}.
\end{align} 
The integral (\ref{wf}) now reads
\begin{widetext}
  \begin{align}
    \int_0^{\infty} d \beta f_n^{\mathrm{inv\chi^2}}\left(\beta ;
    \beta_0\right) w_\mathrm{MB}(K, \beta)
    &= \frac{\beta_0}{\Gamma(n/2)} \left(\frac{n \beta_0}{2}\right)^{n/2}
      \left(\frac{\hbar^2}{2 \pi m}\right)^{1/2} 
      \int_0^{\infty} d \beta \beta^{-(\frac{n+3}{2})}
      \exp\left[-\left(\frac{\beta \hbar^2 K^2}{2m}+
      \frac{n \beta_0}{2 \beta}\right)\right]
      \nonumber \\
    &= \frac{2^{-(1+n/2) }n^{n/2}}{\Gamma(n/2)}
      \left(\frac{\beta_0 \hbar^2}{2 \pi m}\right)^{1/2}  
      \left(\frac{ \beta_0 \hbar^2 K^2}{ n m} \right)^{\frac{1+n}{4}}
      {\mathcal K}_{\frac{1+n}{2}}\left(\sqrt{n \beta_0 \hbar^2 K^2/m} \right),
      \label{Eq57}
  \end{align}
\end{widetext}
where $\mathcal{K}_{\alpha}(x)$ is the modified Bessel function of the
second kind. Note that the inverse $\chi^2$ distribution
(\ref{superstatistics2}) arises naturally when the inverse temperature
$\beta$ is governed by the inverse of an additive stochastic process,
i.e., when $1/\beta$ results from the superposition of many
independent random contributions~\cite{Beck2005}.  In this case,
$\beta$ can be interpreted as the inverse of a sum of squared Gaussian
variables, leading to an inverse $\chi^2$ distribution.  The
corresponding energy distributions in Eq.~(\ref{Eq57}) then exhibits
exponential decay in the tails. Such an exponential behavior has been
documented in a variety of nonequilibrium systems, including vortex
glasses and liquids~\cite{Brito2003}, fusion plasmas~\cite{Sattin},
and harmonic oscillators coupled to solvent environments
\cite{Dixit2013}.

For the present superstatistics, the reactivity gain
$G_n^{\mathrm{inv\chi^2}}$ with respect to a MB distribution at the
same inverse temperature $\beta_0$ is given by
\begin{align}
  G_n^{\mathrm{inv\chi^2}}
  =\frac{2^{-(1+n/2)}n^{n/2}}{\Gamma(n/2)}\,
  \mathcal{G}_{\beta_0}\!\left[\left(\frac{2x}{n}\right)^{\frac{n+1}{4}}
  \!\!\!\!\!
  \mathcal{K}_{\frac{n+1}{2}}\!\left(\sqrt{2nx}\right)\right].
\end{align}

\subsection{Type III: Log-normal superstatistics}

Finally, we consider the log-normal class, which naturally arises when
the inverse temperature results from multiplicative random
processes. In this case,
\begin{align}
  f_s(\beta)=\frac{1}{\sqrt{2 \pi} s \beta}
  \exp\{{-[\ln (\beta/\mu)}]^2/(2 s^2)\},
\end{align}
where $\beta_0=\mu e^{s^2 / 2}$. The variance is
\begin{align}
  \Delta \beta^2=\langle \beta^2 \rangle - \langle \beta \rangle^2
  =  (e^{s^2}-1) \beta_0^2,
  \label{VarianceLog}
\end{align}
and therefore
\begin{align}
  q= \frac{\langle \beta^2 \rangle}{\beta_0^2} =e^{s^2}.
\end{align}
Equilibrium statistical mechanics is recovered {in the limit}
$s \to 0$. {The} log-normal distribution arises naturally when the
inverse temperature $\beta$ is governed by a multiplicative stochastic
process~\cite{Beck2005}. In this case, $\ln \beta$ results from the
superposition of many independent random contributions {and, by virtue
  of the central limit theorem, becomes} Gaussian
distributed. {Consequently,} $\beta$ follows a log-normal
distribution. The {corresponding superstatistical} integral (\ref{wf})
cannot be evaluated in closed form and must therefore be computed
numerically. Log-normal superstatistics is supported by empirical
evidence from systems in which multiplicative and cascade processes,
such as turbulence, play a central role~\cite{stur2,stur3}. It has
also been successfully employed to describe distributions observed in
space plasmas~\cite{Ourabah1} and gravitational systems
\cite{OurabahPRE}, {as well as in} a wide range of complex systems
beyond physics~\cite{nature,Xu}.

For type III, with $y:=\beta/\beta_0$, the reactivity gain
$G_s^{\mathrm{Ln}}$ relative to the MB distribution at the same
$\beta_0$ reads
\begin{equation}
  G_s^{\mathrm{Ln}} =
  \frac{1}{\sqrt{2\pi}\,s}\,\mathcal{G}_{\beta_0}\!\left[
    \int_0^{\infty}\frac{dy}{\sqrt{y}}\,
    e^{-xy-(\ln y+s^2/2)^2/(2s^2)}
  \right].
\end{equation}

\begin{figure*}
  \includegraphics[width=0.45\textwidth, clip=true]{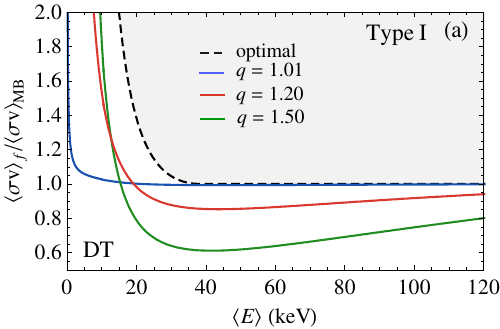}
  \includegraphics[width=0.45\textwidth, clip=true]{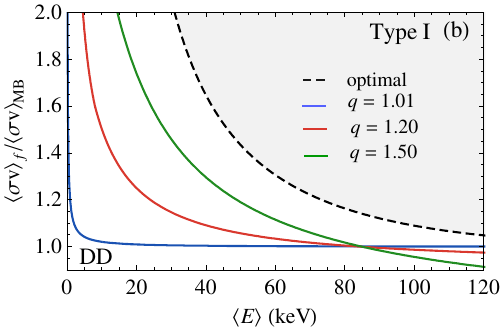}
  \includegraphics[width=0.45\textwidth, clip=true]{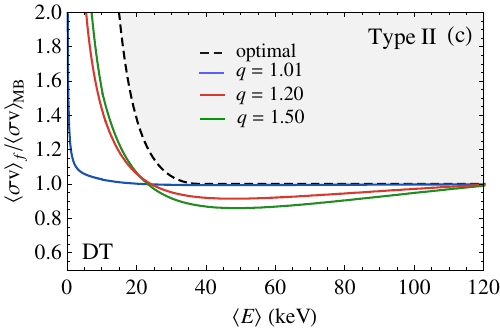}
  \includegraphics[width=0.45\textwidth, clip=true]{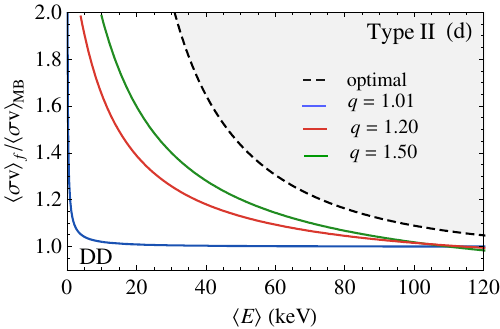}
  \includegraphics[width=0.45\textwidth, clip=true]{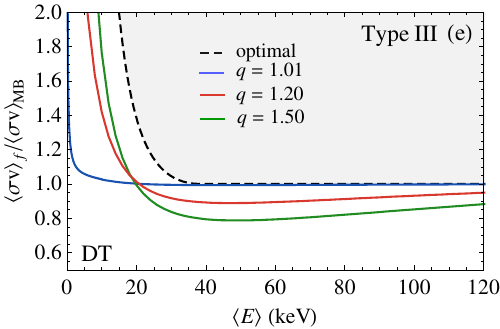}
  \includegraphics[width=0.45\textwidth, clip=true]{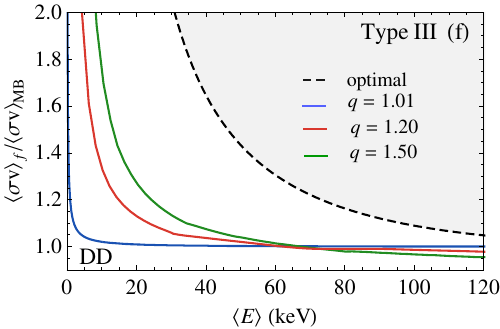}
  \caption{Gain factor versus the average energy $\langle E \rangle$
    for the three representative superstatistics, type I (top plots),
    type II (middle), and type III (bottom), each shown for three
    different values of $q$, and the exclusion region (in gray) for the
    optimal gain according to the concave envelope, delimited by the
    dashed line.  The cases on the left are relative to DT fusion; the
    cases on the right are relative to DD fusion.}
  \label{OptimalFig3}
\end{figure*}

Note that, in order to assess whether a given superstatistical
distribution enhances the reactivity, the comparison should be
performed at the same available kinetic energy $\langle E \rangle$,
rather than at the same value of $\beta_0$. For the three classes
considered here, the average energy reads
\begin{align}\label{energy}
  \begin{aligned}
    \langle E \rangle_{}
    =&
       \left (\frac{n}{n-2} \right )\frac{1}{2\beta_0} \quad \text{(type I)},
    \\
    \langle E \rangle_{}
    =&
       \left(\frac{n+2}{n}\right)\frac{1}{2\beta_0}  \quad \text{(type II)},
    \\
    \langle E \rangle_{}
    =&
       \frac{e^{s^2}}{2\beta_0}  \quad \text{(type III)}.   
  \end{aligned}
\end{align}
The resulting equal-energy reactivities can then be expressed as
\begin{align}
  \langle \sigma v \rangle_f= \left(\frac{\hbar^2}{2\pi m}\right)^{1 / 2}
  \int_{-\infty}^{+\infty} d k \int_{-\infty}^{+\infty} d K \quad I_f(k,K). 
  \label{IntegralkK}
\end{align}
The functions $I_f(k,K)$ corresponding to the different
superstatistical classes, written in terms of the common parameter
$q=\langle \beta^2 \rangle / \beta_0^2$, are summarized in Table~I.

\begin{figure*}
  \includegraphics[width=0.48\textwidth, clip=true]{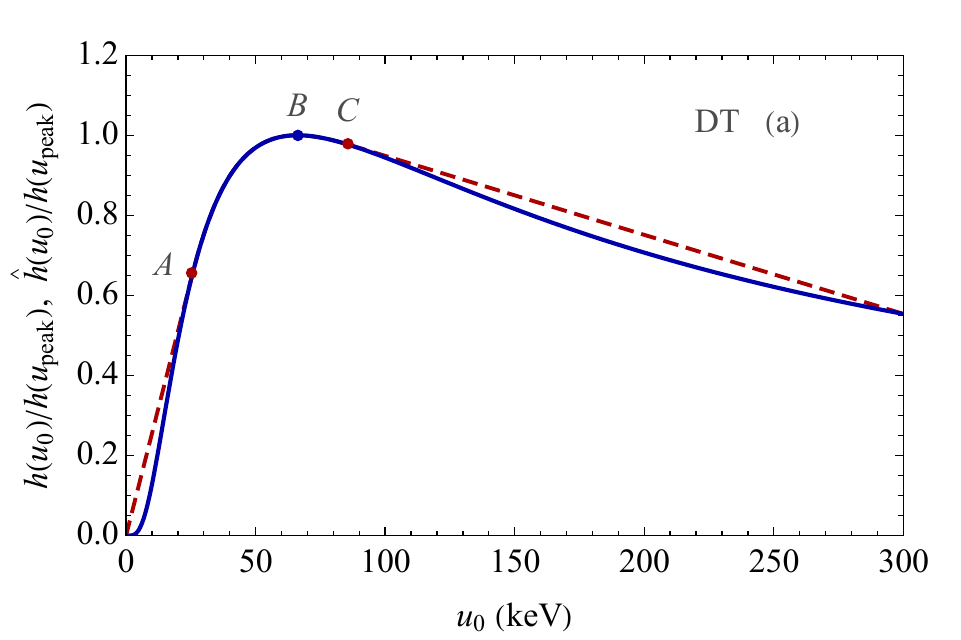}
  \includegraphics[width=0.48\textwidth, clip=true]{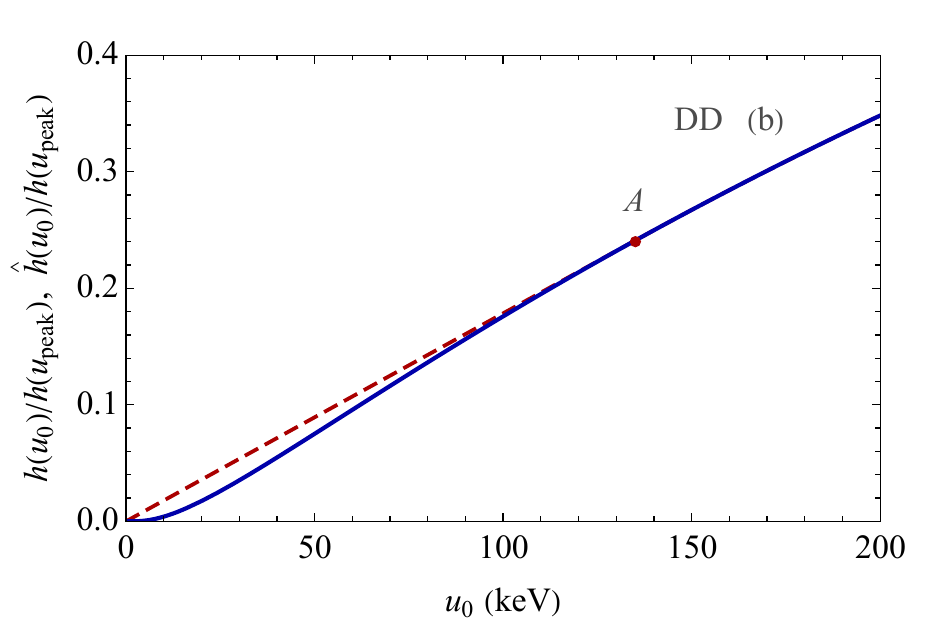}
  \includegraphics[width=0.48\textwidth, clip=true]{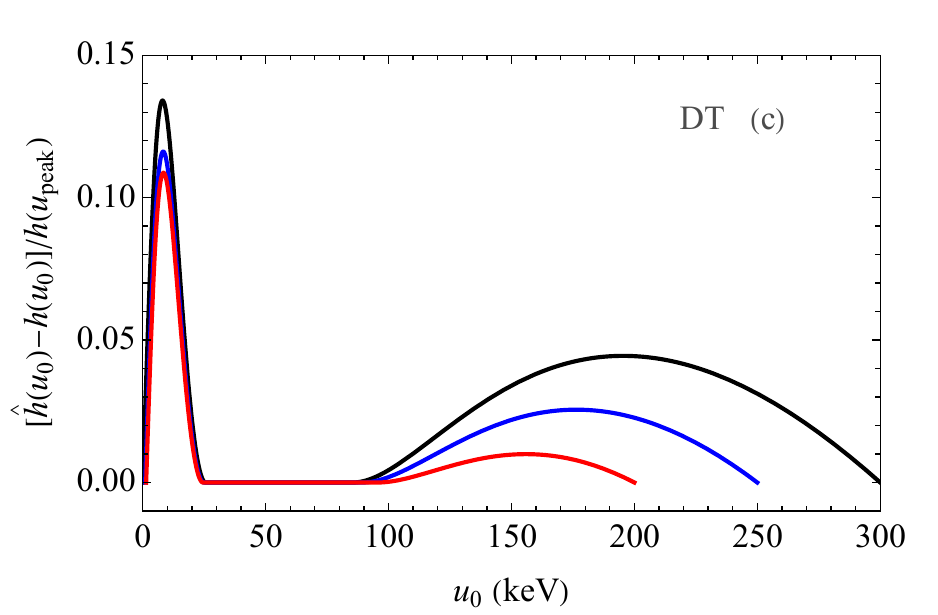}
  \includegraphics[width=0.48\textwidth, clip=true]{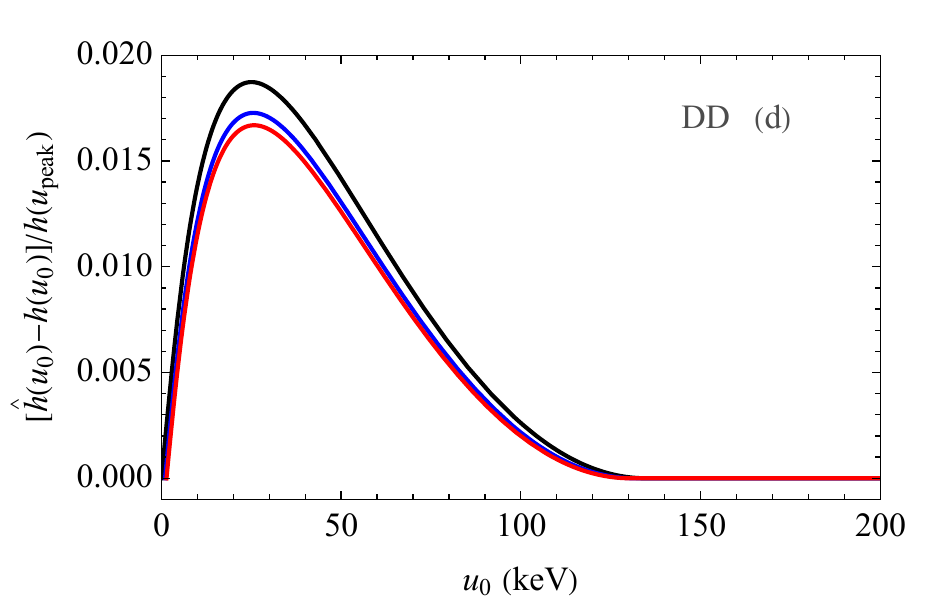}
  \caption{Function $h(u_0)$ representing the empirical reactivity
    calculated from \cite{Bosch} (blue solid line) and its concave
    envelope $\hat{h}(u_0)$ (red dashed line) for (a) DT fusion and
    (b) DD fusion. In both cases, the functions are normalized to the
    peak value of $h(u)$ achieved at point $B$:
    $u_{\text{peak}} = 66.157\text{ keV}$ for DT and
    $u_{\text{peak}} = 1.216\text{ MeV}$ for DD (point not shown).  In
    (a), the boundary points of the left and right regions where
    $\hat{h}(u_0) > h(u_0)$ are
    $u_1^{(l)} = u_{\text{min}} = 0\text{ keV}$,
    $u_2^{(l)} = 25.652\text{ keV}$ (point $A$),
    $u_1^{(r)} = 85.386\text{ keV}$ (point $C$), and
    $u_2^{(r)} = u_{\text{max}} = 300\text{ keV}$. Conversely, (b)
    shows only the left region with
    $u_1^{(l)} = u_{\text{min}} = 0\text{ keV}$ and
    $u_2^{(l)} = 134.843\text{ keV}$ (point $A$).  Peak-normalized
    reactivity gain $[\hat{h}(u_0) - h(u_0)]/h(u_{\text{peak}})$ as a
    function of $u_0$ when using the optimized energy distribution for
    (c) DT and (d) DD. Three curves are shown in each panel,
    corresponding to different choices of $u_{\text{min}}$ and
    $u_{\text{max}}$ for DT (black: $0, 300$; blue: $1, 250$; red:
    $1.4, 200$) and $u_{\text{min}}$ for DD (black: $0$; blue: $1$;
    red: $1.4$); all values are in keV.}
  \label{OptimalFig4}
\end{figure*}

From Eq.~(\ref{energy}), it is useful to introduce an effective
temperature $T_{\mathrm{eff}}$, defined as the temperature a MB
distribution would require to have the same average kinetic energy as
the corresponding superstatistical distribution. This quantity is
shown in Fig.~\ref{OptimalFig2}(b) as a function of
the common parameter $q$.  Figure~\ref{OptimalFig2} shows that, for the three
superstatistical classes, the effective temperature exceeds that of
the corresponding MB distribution. Moreover, a clear hierarchy
emerges. For the same value of $q$, type I exhibits the largest
deviation from the Maxwellian reference, whereas type II remains the
closest to it. This behavior is consistent with the trends observed in
Fig.~\ref{OptimalFig1}.

Finally, it should be noted that the power-law asymptotic behavior of
the type I distribution imposes restrictions on the parameter $n$, or
equivalently on $q$. Requiring a finite average energy
$\langle E \rangle$ (or a finite effective temperature
$T_{\mathrm{eff}}$) leads to $n>2$, which is equivalent to
$q<2$. In addition, in the physically relevant three-dimensional case,
quite general arguments suggest the stronger bound $q\lesssim 1.66$,
which appears to be compatible with available real-world data
~\cite{Our2025}.

\section{Reactivity gain with empirical formulas for fusion cross
  sections}

As an application of the considerations described in the previous two
sections, here, we consider the realistic case of DT and DD fusion
reactivities. Following Ref.~\cite{Bosch}, the cross sections of these
fusion processes are parametrized as
\begin{align}
  \sigma=S(E) \frac{e^{-B_G/\sqrt{E}}}{E},
\end{align}
where $B_G= \pi \alpha Z_1 Z_2 \sqrt{2 m c^2}$ is the Gamow constant,
with $\alpha$ being the fine structure constant, $m$ being the reduced
mass, and $Z_1$ and $Z_2$ being the electric charges of the nuclei in
elementary charge units, and $S(E)$ is the so-called astrophysical $S$
function, expressed in terms of a Pad\'e polynomial as
\begin{align}
  S(E)=\frac{A_1+E(A_2+E(A_3+E(A_4+EA_5)))}{1+E(B_1+E(B_2 +E(B_3+EB_4)))},
\end{align}
with the values of $B_G$, $A_i$, $B_j$, with $i=1,\dots,5$ and
$j=1,\dots,4$, being reported in Table IV of Ref.~\cite{Bosch} for
various energy ranges depending on the specific fusion reaction. The
above parameters are specified for both DT and DD fusion. For the DD
case, the total reactivity includes the contributions from both
reaction channels.

A first analysis consists of comparing the optimal reactivity
discussed in Sec. II and the reactivities arising from the
superstatistics in Sec. III, both using the empirical cross sections
of Ref.~\cite{Bosch}.  Figure~\ref{OptimalFig3} shows the reactivity
gain obtained using the three superstatistics described in Sec. III
(top, middle, and bottom rows) relative to the MB distribution at the
same average energy, for DT fusion (left column) and DD fusion (right
column).  All cases show an enhancement of reactivity at low average
energies, confirming earlier results reported in
Ref.~\cite{Squarer}. In the case of DT fusion, such an enhancement at
low average energy is followed by a suppression of reactivity at
higher average energies, especially in the case of superstatistics of
type I.  In each panel of Fig.~\ref{OptimalFig3}, the shaded region
indicates the exclusion zone for the reactivity gain determined by the
optimal reactivity in Sec. II, with its boundary represented by a
dashed line. As expected, we find that the gain of each considered
superstatistics is always upper bounded by the corresponding shaded
contour.  The distance of the superstatistics curves from the dashed
line indicates that considerable gain can be further achieved by using
the optimal reactivity.
  
The gain obtained from using optimized reactivities depends on the
average energy; specifically, this gain occurs in the regions where
the MB reactivity is exceeded by its concave envelope.  As discussed
in Sec. II, there are two such regions—one at low energy and the other
at high energy [see Figs.~\ref{OptimalFig4}(a) and
\ref{OptimalFig4}(b) for the DT and DD fusion reactions]—and their
extensions depend on the choice of $u_{\mathrm{min}}$ and
$u_{\mathrm{max}}$, which define the boundaries of the supported MB
distribution.

It is evident [see Figs.~\ref{OptimalFig4}(c) and
\ref{OptimalFig4}(d)] that the gain at low energies increases as
$u_{\mathrm{min}}$ decreases, whereas at high energies, larger gains
are achieved by increasing $u_{\mathrm{max}}$.  Indeed, decreasing
$u_{\mathrm{min}}$ or increasing $u_{\mathrm{max}}$ shifts the
matching points $u_2^{(l)}$ and $u_1^{(r)}$ closer to
$u_\mathrm{peak}$, causing the concave envelope to progressively
exceed the MB reactivity.  However, the actual gain in using this
maximum span can be limited. First, the case of average energy close
to 0 keV is not of physical or engineering interest because the
absolute reactivity tends to zero in that limit anyway, and the
measurements are limited to non zero values of energy.  Second, the
maximum value of $u_{\mathrm{max}}$ is limited by the actual
experimental data taken at the highest temperatures and can also be
limited by the opening of new environmental conditions, for instance
the onset of an electron-positron plasma at temperatures of the order
of 0.5--1 MeV.

The advantage of using an energy distribution based on the concave
envelope with respect to a MB distribution with the same average
energy can be evaluated for DT and DD reactions as follows. Regarding
the DT fusion case, the peak reactivity is found to be
$h(u_{\mathrm{peak}}) = 8.9 \times 10^{-16} \, \mathrm{cm^3 s^{-1}}$,
using the parametrization from Ref.~\cite{Bosch}. At the reference
energy of $u_0 = 10$ keV, we find a maximum reactivity difference of
$\hat{h}(u_0) - h(u_0) = 1.2 \times 10^{-16} \, \mathrm{cm^3 s^{-1}}$,
as shown in Fig.~\ref{OptimalFig4}(c) where
$[\hat{h}(u_0) - h(u_0)]/h(u_{\text{peak}}) \simeq 0.135$ in the most
favorable case. Given
$h(u_0) = 1.15 \times 10^{-16} \, \mathrm{cm^3 s^{-1}}$, this
corresponds to a reactivity gain of $\hat{h}(u_0)/h(u_0) = 2.04$ when
we use the optimized energy distribution compared to a MB distribution
with the same average energy $u_0$. This optimized distribution is a
bimodal MB energy distribution with $w_1 = 0.61$, as derived from
Eq.~(\ref{w1}). Repeating the calculation for DD fusion, we obtain a
maximum reactivity gain of 1.7 at $u_0 = 23 \, \mathrm{keV}$,
utilizing a bimodal MB energy distribution with $w_1 = 0.84$.

It is worth mentioning that this approximately twofold gain could take
place in an energy region of immediate interest for magnetically
confined plasmas. From a practical standpoint, the protocol will
consist of using for the DT reaction, instead of a MB-distributed
plasma at an average energy of $u_0=10$ keV, a bimodal MB energy
distribution made of a very cold plasma, at a percentage of
$\simeq 60 \%$, and a hot component at $u_0 \sim 25$ keV.  Bimodal MB
distributions were already achieved at the Joint European Torus (JET)
as reported in Ref.~\cite{Bhatnagar} and more recently discussed in
detail~\cite{Maslov}.

\section{Conclusions}
We discussed the optimization of fusion reactivity with respect to the
energy distribution of the reacting ions, confirming the preliminary
findings reported in Ref.~\cite{Squarer} regarding the role of the
concavity of the reactivity curve as a function of the average
particle energy. This result is now established within the more
rigorous framework of a constrained functional optimization problem,
thereby increasing in clarity and generality. As long as the average
energy remains within a range where the reactivity curve is concave,
the MB energy distribution optimizes the reactivity. Outside this
interval, the optimal solution is generally obtained by choosing a
bimodal MB distribution, with temperatures determined by the straight
lines that tangentially connect the boundary points of the two
regions.  While the three discussed classes of superstatistics can
exceed the reactivity of their corresponding MB energy distributions,
they all remain below the optimization boundary of the reactivity
curve. Finally, we conjectured a relationship between optimized energy
distributions and superstatistics. Most of these considerations on
reactivity optimization can be extended to a more general setting,
such as reactions of chemical and biological
interest~\cite{Bartlova,Wild,BeckTsallis}.

Regarding their applicability to confined plasmas, optimized energy
distributions could play a significant role in an interesting regime
centered around $u_0\approx 10$ keV for DT fusion and $u_0\approx 30$
keV for DD fusion. In these regions, one can expect peak-normalized
reactivity gains $[\hat{h}(u_0) - h(u_0)]/h(u_{\text{peak}})$ of
approximately 10\% for DT and 1.5\% for DD, see
Figs.~\ref{OptimalFig4}(c) and \ref{OptimalFig4}(d), which translate
into a boost in reactivity gain up to a factor of about 2 for both DD
and DT mixtures. This requires envisaging demonstrations of
engineered bimodal MB energy distributions.  Within the context of a
much simpler confinement geometry, such as a harmonic trap, numerical
simulations have demonstrated that both bimodal MB~\cite{Bala10} and
$\kappa$ energy distributions~\cite{Bala12} are generated during a
thermalization transient. In the former case, a mechanism was proposed
where bimodal distributions arise from a local mismatch in heat
capacities between the two species undergoing thermalization. A
compelling future direction would be to extend these considerations to
the more realistic confinement scenarios of tokamaks, leveraging
existing codes designed to identify non-equilibrium plasma
features~\cite{ATEP1,ATEP2}.

\begin{acknowledgements}
  We thank Lorenzo Bertini for a fruitful discussion on
  the solution of constrained linear variational problems.
\end{acknowledgements}


\begin{thebibliography}{99}

\bibitem{ITER1} P.-H. Rebut, ITER: the
  first experimental fusion reactor, Sci. Direct
  \textbf{30}, 85 (1995).

\bibitem{ITER2} V. Mukhovatov, \textit{et al.}, Overview of physics
  basis for ITER, Plasma Phys. Control. Fusion \textbf{45}, A235-A252
  (2003).

\bibitem{Onofrio} R. Onofrio, Concepts for a Deuterium–Deuterium
  fusion reactor, J. Exp. Theor. Phys. \textbf{127}, 883 (2018).

\bibitem{OnofrioPresilla} R. Onofrio and C. Presilla, State dependence
  of tunneling processes and thermonuclear fusion, Nucl. Phys A
  \textbf{1043}, 122830 (2024).

\bibitem{GWS} B.~C.~Lutfu\u{g}lu, F.~Akdeniz, and O.~Bayrak,
  Scattering, bound, and quasi-bound states of the generalized
  symmetric Woods-Saxon potential, J.~Math.~Phys. \textbf{57}, 032103
  (2016).

\bibitem{Hwang} E. Hwang, H. Ko, K. Heo, M.-K. Chooun, and D. Jang,
  Reinvestigating the Gamow factor of reactions on light nuclei,
  Astr. J. \textbf{955}:79 (2023).

\bibitem{Harvey} R. W. Harvey, M. G. McCoy, G. D. Kerbel, and
  S. C. Chiu, ICRF fusion reactivity enhancement in tokamaks,
  Nucl. Fusion \textbf{26}, 43 (1986).

\bibitem{Nath} D. Nath, R. Majumdar, and M. S. Kalra, Thermonuclear
  Fusion Reactivities for Drifting Tri-Maxwellian Ion Velocity
  Distributions, J. Fusion Energy \textbf{32}, 457 (2013).

\bibitem{Majumdar} R. Majumdar and D. Das, Estimation of total fusion
  reactivity and contribution from supra-thermal tail using
  3-parameter Dagum ion speed distribution, Annals Nucl. Energy
  \textbf{97}, 66 (2016).

\bibitem{Kolmes} E. J. Kolmes, M. E. Mlodik, and N. J. Fisch, Fusion
  yield of plasma with velocity-space anisotropy at constant energy,
  Physics of Plasmas \textbf{28}, 052107 (2021).

\bibitem[Li et al.(2022)]{Li} K. Li, Z. Y. Liu, Y. L. Yao, Z. H. Zhao,
  C. Dong, D. Li, S. P. Zhu, X. T. He, and B. Qiao, Modification of
  the fusion energy gain factor in magnetic confinement fusion due to
  plasma temperature anisotropy, Nucl. Fusion \textbf{62}, 086026
  (2022).

\bibitem{Xie1} H. Xie, A simple and fast approach for computing the
  fusion reactivities with arbitrary ion velocity distributions,
  Comput. Phys. Commun. \textbf{292}, 108862 (2023).

\bibitem{Xie2} H. Xie, M. Tan, D. Luo, Z. Li and B. Liu, Fusion
  reactivities with drift bi-Maxwellian ion velocity distributions,
  Plasma Phys. Control. Fusion \textbf{65}, 055019 (2023).

\bibitem{Kong} H. Kong, H. Xie, B. Liu, M. Tan, D. Luo, Z. Li, and
  J. Sun, Enhancement of fusion reacivity under non-Maxwellian
  distributions: Effects of drift-ring-beam, slowing-down, and Kappa
  super-thermal distributions, Plasma Phys. Control. Fusion
  \textbf{66}, 015009 (2024).

\bibitem{Squarer} B. I. Squarer, C. Presilla, and R. Onofrio,
  Enhancement of fusion reactivities using non-Maxwellian energy
  distributions, Phys. Rev. E \textbf{109}, 025207 (2024).

\bibitem{Coppi} B. Coppi and B. Basu, Non-thermal fusion burning
  processes, relevant collective modes and gained perspectives,
  Nucl. Fusion \textbf{64}, 126037 (2024).

\bibitem{Ye} Y. Ye, W. Zhang, and B. Wan, Enhancement of the fusion
  reactivity due to the D-T non-Maxwellian ion distribution and its
  impact on Lawson criterion, Phys. Plasmas \textbf{32}, 092504
  (2025).

\bibitem{Nicolas} T. Nicolas, Dependence on temperature of isotropic
  non-thermal fusion reactivity enhancements of the DT, D${}^3$He and
  DD reactions, Plasma Phys. Control. Fusion \textbf{68}, 025011
  (2026).

\bibitem{Rider} T. H. Rider, Fundamental limitation on plasma fusion
  systems not in thermodynamics equilibrium, Phys. Plasmas \textbf{4},
  1039 (1997).

\bibitem{Livadiotis1} G. Livadiotis and D. J. McComas, Invariant kappa
  distributions in space plasmas out of equilibrium,
  Astrophys. Journ. \textbf{741}, 1 (2011).

\bibitem{Tsallis} C. Tsallis, {\sl Introduction to Nonextensive
    Statistical Mechanics}, vol. 34 (Springer, 2009).

\bibitem{Livadiotis2} G. Livadiotis, Thermodynamic origin of kappa
  distributions, EPL \textbf{122}, 50001 (2018).

\bibitem{Sattin} F. Sattin and L. Salasnich, Multiparameter
  generalization of nonextensive statistical mechanics, Phys. Rev. E
  \textbf{65}, 035106(R) (2002).

\bibitem{stur2} A. Reynolds, Superstatistical Mechanics of
  Tracer-Particle Motions in Turbulence, Phys. Rev. Lett. \textbf{91},
  084503 (2003).

\bibitem{stur3} S. Jung and H. L. Swinney, Velocity difference
  statistics in turbulence, Phys. Rev. E \textbf{72}, 026304 (2005).

\bibitem{Wilk} G. Wilk and Z. Wlodarczyk, Consequences of temperature
  fluctuations in observables measured in high-energy collisions,
  Eur. Phys. J. A \textbf{48}, 161 (2012).

\bibitem{Ourabah1} K. Ourabah, Demystifying the success of empirical
  distributions in space plasmas, Phys. Rev. Res. \textbf{2}, 023121
  (2020).

\bibitem{OurabahPRE} K. Ourabah, Generalized statistical mechanics of
  stellar systems, Phys. Rev. E \textbf{105}, 064108 (2022).

\bibitem{Market} M. Denys, T. Gubiec, R. Kutner, M. Jagielski, and
  H. E. Stanley, Universality of market superstatistics, Phys. Rev. E
  \textbf{94}, 042305 (2016).

\bibitem{air} H. He, B. Sch\"{a}fer, and C. Beck, Spatial
  heterogeneity of air pollution statistics in Europe,
  Sci. Rep. \textbf{12}, 12215 (2022).

\bibitem{nature} B. Sch\"{a}fer, C. Beck, K. Aihara, D. Witthaut, and
  M. Timme, Non-Gaussian power grid frequency fluctuations
  characterized by Lévy-stable laws and superstatistics, Nat. Energy
  \textbf{3}, 119 (2018).

\bibitem{Jensen} J. L. W. V. Jensen, Sur les fonctions convexes et les
  inégalités entre les valeurs moyennes, Acta Mathematica \textbf{30},
  175 (1906).

\bibitem{Bosch} H.-S. Bosch and G. M. Hale, Improved formulas for
  fusion cross section and thermal reactivities, Nucl. Fusion
  \textbf{32}, 611 (1992).

\bibitem{Satchler} G. R. Satchler, {\sl Introduction to Nuclear
    Reactions} (Palgrave, Macmillan, London,1990).

\bibitem{Alterman} S. Alterman, J. Choi, R. Durst, S. M. Fleming, and
  W. K. Wootters, The Boltzmann distribution and the quantum-classical
  correspondence, J. Phys. A: Math Theor. \textbf{51}, 345301 (2018).

\bibitem{Chenu} A. Chenu and M. Combescot, Many-body formalism for
  thermally excited wave packets: A way to connect the quantum regime
  to the classical regime, Phys. Rev. A \textbf{95}, 062124 (2017).

\bibitem{GMP} S. Karlin and W. .J. Studden, {\sl Tchebycheff Systems:
    With Applications in Analysis and Statistics}(Interscience
  Publishers, New York,1966).

\bibitem{SBE1} A.~Madansky, Bounds on the Expectation of a Convex
  Function of a Multivariate Random Variable, The Annals of
  Mathematical Statistics \textbf{30}, 743 (1959).

\bibitem{SBE2} H.~P.~Edmundson, Bounds on the Expectation of a Convex
  Function of a Random Variable, RAND Corporation Papers, P982 (1957).

\bibitem{Cara} C.~Carath\'eodory, \"Uber den Variabilit\"atsbereich
  der Koeffizienten von Potenzreihen, die gegebene Werte nicht
  annehmen, Mathematische Annalen, \textbf{64}, 95–115 (1907).

\bibitem{Beck1} C. Beck, Dynamical Foundations of Nonextensive
  Statistical Mechanics, Phys. Rev. Lett. \textbf{87}, 180601 (2001).

\bibitem{Beck2} C. Beck and E. G. D. Cohen, Superstatistics, Physica A
  \textbf{322}, 267 (2003).

\bibitem{Ourabah2} K. Ourabah, Reaction rates in quasiequilibrium
  states, Phys. Rev. E \textbf{111}, 034115 (2025).

\bibitem{Beck2005} C. Beck, E. G. D. Cohen, and H. L. Swinney, From
  time series to superstatistics, Phys. Rev. E \textbf{72}, 056133
  (2005).

\bibitem{Vander2008} E. Van der Straeten and C. Beck, Superstatistical
  distributions from a maximum entropy principle, Phys. Rev. E
  \textbf{78}, 051101(2008).

\bibitem{azerty5} G. Livadiotis and D. J. McComas, Beyond kappa
  distributions: Exploiting Tsallis statistical mechanics in space
  plasmas, J. Geophys. Res. \textbf{114}, A11105 (2009).

\bibitem{plasma1} B. Liu and J. Goree, Superdiffusion and Non-Gaussian
  Statistics in a Driven-Dissipative 2D Dusty Plasma,
  Phys. Rev. Lett. \textbf{100}, 055003 (2008).

\bibitem{Astro1} J. C. Carvalho, J. D. do Nascimento, Jr., R. Silva,
  and J. R. De Medeiros, Non-Gaussian statistics and stellar
  rotational velocities of main-sequence field,
  Astrophys. J. \textbf{696}, L48 (2009).

\bibitem{galaxy} A. L. B. Ribeiro, P. A. A. Lopes, and M. Trevisan,
  Non-Gaussian velocity distributions — the effect on virial mass
  estimates of galaxy groups, MNRAS \textbf{413}, L81 (2011).
\bibitem{Our2020} K. Ourabah, Quasiequilibrium self-gravitating
  systems, Phys. Rev. D \textbf{102}, 043017 (2020).

\bibitem{Mahsa} M. Iranmanesh, H. A. Kermani, and K. Ourabah,
  Superstatistics and stellar rotation: Modeling velocity
  distributions in six stellar groups, New Astronomy \textbf{119},
  102408 (2025).

\bibitem{cold1} E. Lutz, Anomalous diffusion and Tsallis statistics in
  an optical lattice, Phys. Rev. A. \textbf{67}, 051402(R) (2003).

\bibitem{cold2} P. Douglas, S. Bergamini, and F. Renzoni, Tunable
  Tsallis Distributions in Dissipative Optical Lattices,
  Phys. Rev. Lett. \textbf{96}, 110601 (2006).
 
\bibitem{cold3} G. B. Bagci and U. Tirnakli, Self-organization in
  dissipative optical lattices, Chaos \textbf{19}, 033113 (2009).

\bibitem{HE1} V. Khachatryan \textit{et al.} (CMS Collaboration),
  Transverse-Momentum and Pseudorapidity Distributions of Charged
  Hadrons in $pp$ Collisions at $\sqrt{s}=7$ TeV,
  Phys. Rev. Lett. \textbf{105}, 022002 (2010).
 
\bibitem{HE2} A. Adare \textit{et al.} (PHENIX Collaboration),
  Measurement of neutral mesons in $p+p$ collisions at $\sqrt{s}=200$
  GeV and scaling properties of hadron production, Phys. Rev. D
  \textbf{83}, 052004 (2011).

\bibitem{Brito2003} C. Brito, I. S. Aranson, and H. Chaté, Vortex
  glass and vortex liquid in oscillatory media,
  Phys. Rev. Lett. \textbf{90}, 068301 (2003).

\bibitem{Dixit2013} P. D. Dixit, A maximum entropy thermodynamics of
  small systems, J. Chem. Phys. \textbf{138}, 184111 (2013).

\bibitem{Xu} D. Xu and C. Beck, Transition from lognormal to
  $\chi^2$-superstatistics for financial time series, Physica A
  \textbf{453}, 173 (2016).

\bibitem{Our2025} K. Ourabah, Superstatistics in the context of
  relativity, Phys. Rev. Res. \textbf{7}, 033190 (2025).

\bibitem{Bhatnagar} V. P. Bhatnagar, J. Jacquinto, D.F.H. Start, and
  B.J.D. Tubbing, High concentration minority ion cyclotron resonance
  heating in JET, Nucl. Fusion \textbf{33}, 83 (1993).

\bibitem{Maslov} M. Maslov, E. Lerche, F. Auriemma, E. Belli,
  C. Bourdelle, C. Challis, A. Chomiczewska, A. Dal Molin,
  J. Eriksson, J. Garcia et al., JET D-T scenario with optimized
  non-thermal fusion, Nucl. Fusion \textbf{63}, 112002 (2023).

\bibitem{Bartlova} M. Bartlov\'a, The influence of nonequilibrium
  distribution function on values of reaction rate constants,
  Czechoslovak Journal of Physics \textbf{50}, 289 (2000).

\bibitem{Wild} R. Wild, M. N\"otzold, M. Simpson, T. D. Tran, and
  R. Wester, Tunneling measured in a very slow ion-molecule reaction,
  Nature \textbf{615}, 425 (2023).

\bibitem{BeckTsallis} C. Beck and C. Tsallis, Anomalous velocity
  distributions in slow quantum-tunneling chemical reactions,
  Phys. Rev. Res. \textbf{7}, L01081 (2025).

\bibitem{Bala10} R. Onofrio and B. Sundaram, Generation of
  non-Maxwell-Boltzmann energy distributions during thermalization of
  interacting baths, Phys. Rev. E \textbf{111}, 064127 (2025).

\bibitem{Bala12} R. Onofrio and B. Sundaram, Dynamics of
  thermalization in classical three-dimensional many-body systems:
  Non-Maxwell-Boltzmann distributions and the role of anisotropic
  trapping, Phys. Rev. E \textbf{112}, 044148 (2025).

\bibitem{ATEP1} Ph. Lauber, M. Falessi, G. Meng, T. Hayward-Schneider,
  V.-A. Popa, F. Zonca, and M. Schneider, ATEP: an advanced transport
  model for energetic particles, Nucl. Fusion \textbf{64}, 096010
  (2024).

\bibitem{ATEP2} G. Meng, Ph. Lauber, Z. Lu, A. Bergmann, and
  M. Schneider, Energetic particles transport in constants of motion
  space due to collisions in tokamak plasmas, Nucl. Fusion
  \textbf{64}, 096009 (2024).
 
\end{thebibliography}
\end{document}